\documentclass[aps,prb,twocolumn,reprint,titlepage,longbibliography]{revtex4-1}

\usepackage{graphicx}
\usepackage{color}
\usepackage{dcolumn}
\usepackage{amsfonts}
\usepackage{bm}
\usepackage[utf8]{inputenc}

\usepackage{epstopdf}

\begin{document}

\title{Bidirectional flow of two-dimensional dusty plasma under asymmetric periodic substrates driven by unbiased external excitations}
\author{Wei Li$^1$$^,$$^2$ $^\ast$, D. Huang$^2$, C. Reichhardt$^3$, C. J. O. Reichhardt$^3$, and Yan Feng$^2$ $^\ast$}
\affiliation{
$^1$ School of Science, Nantong University, Nantong 226019, China\\
$^2$ Jiangsu Key Laboratory of Thin Films and Institute of Plasma Physics and Technology, School of Physical Science and Technology, Soochow University, Suzhou 215006, China\\
$^3$ Theoretical Division, Los Alamos National Laboratory, Los Alamos, New Mexico 87545, USA\\
$\ast$ E-mail: liwei@ntu.edu.cn; fengyan@suda.edu.cn}

\date{\today}

\begin{abstract}

Collective transport properties of a one-dimensional asymmetric periodic substrate (1DAPS) modulated two-dimensional dusty plasma (2DDP) driven by an unbiased sinusoidal excitation force are investigated using Langevin dynamical simulations. 
It is discovered that, by changing the amplitude and frequency of the unbiased sinusoidal external excitations, as well as the depth of 1DPAS, both the direction and speed of the persistent particle flow can be adjusted, i.e.,
both the flow rectification and its reversal of the ratchet effect of the steady drift motion for particles are achieved using various excitations.
For the studied 2DDP under the 1DAPS, when the amplitude of the excitation increases from zero, the magnitude of the overall drift velocity increases from zero to its first maximum in the easy direction of the 1DAPS, next decreases gradually back to zero, and then increases from zero to its second maximum in the hard direction of 1DAPS before finally gradually decaying. 
It is found that, as the frequency of the excitation and the depth of 1DAPS change, the maximum overall drift velocity also varies, and the corresponding magnitude of the excitation varies simultaneously.
The observed ratchet effect in both the easy and hard directions of 1DAPS for 2DDP is attributed to the combination of the spatial symmetry breaking of 1DAPS and the inertial effects of particles, which is further confirmed by the three different presented diagnostics.
\end{abstract}

\maketitle

\section{Introduction}

When overdamped particles are modulated by an asymmetric periodic substrate, such as a series of sawtooth shaped
potential wells, the so-called rocking ratchet effect can be exhibited, in which an applied ac drive generates a net dc flow of particles \cite{Reimann02,Arzola11}. 
If an asymmetric periodic substrate is flashed on and off in
a system with thermal fluctuations,
a flashing ratchet effect may still appear, even without an ac drive \cite{Reimann02,Rousselet94}.
Various ratchet effects have been studied in a wide variety of systems, 
including colloidal assemblies \cite{Reimann02,Arzola11,Rousselet94},
biological systems \cite{Lau17},
magnets \cite{PerezJunquera08,Franken12},
cold atom arrays \cite{Jones04},
solid state devices \cite{Roeling11,Grossert16},
quantum systems \cite{Linke99,Salger09},
fluids on asymmetric substrates \cite{Lagubeau11},
superconducting vortices \cite{Lee99},
granular matter \cite{Wambaugh02} and active matter
\cite{Reichhardt17a}.
In these asymmetric substrate modulated systems, there is generally
an easy flow direction for particles
in which the force needed to move over the substrate barrier is smaller than
that for the flow in the
opposite or hard direction. In most ratchet
systems, particle flows occur only in the easy direction;
however, there are still some cases where the dc drift velocity
undergoes a reversal under different conditions, so that particle flows may also
preferentially occur in the hard direction.
There may even be multiple reversals of the flow direction
as a function of the ac driving amplitude, or the frequency, or the particle density
\cite{Derenyi95,Mateos00,deSouzaSilva06a,Gillijns07,Villegas03,Lu07,McDermott16}.
These reversals may arise
due to inertial effects \cite{Villegas03} or collective effects
\cite{Derenyi95}, in which localized quasiparticles
appear or there are multiple effective species that
experience inverted
potentials \cite{Villegas03,Lu07,McDermott16}.

One of the best studied examples of
a particle-based system coupled to a periodic asymmetric substrate
that shows dc current reversals as
a function of ac drive amplitude and particle density is 
vortices in type-II superconductors
\cite{Lee99,Villegas03,deSouzaSilva06a,Lu07,Gillijns07,McDermott16,Reichhardt05,Yu07,Reichhardt10a,PerezdeLara11,Shklovskij14,Reichhardt15}.
The vortices behave like
overdamped particles that have intermediate range interactions similar to the
screened Coulomb interactions.
At higher vortex densities, the system forms a triangular lattice, while at low
vortex densities the system is in the liquidlike state.
In the study of a two-dimensional (2D)
assembly of superconducting vortices at a low density coupled
to an asymmetric quasi-one dimensional (q1D)
sawtooth potential~\cite{Lee99}, it is found that under an ac drive, the vortices move in the
easy flow direction of the substrate.
Similar results
are obtained for vortices in asymmetric channels \cite{Wambaugh02}.
In the experiment performed with
superconducting vortices on asymmetric substrates~\cite{Villegas03}, the vortex motion at low densities is
in the easy direction as expected,
but a reversal of the flow into the hard direction occurs
as the vortex density increases, while for sufficiently large
vortex densities, the dc flow disappears.
The current reversals are attributed to
the collective interactions
between the vortices,
since at high densities the system
can be viewed as containing
two effective species of particles.
The second species interacts with the potential landscape
created by the first species
rather than directly experiencing the substrate asymmetry,
and the asymmetry of this potential landscape 
is inverted as compared with the substrate asymmetry.
In the study of a simpler version of the 2D superconducting vortex
assembly on a q1D asymmetric potential~\cite{Lu07},
it is found that at low densities,
the ratchet motion is in the easy
flow direction, in agreement
with Ref.~\cite{Lee99};
however, at higher drives
there is a reversal in the flow
when two rows of vortices occupy each substrate minimum and undergo
a buckling transition
into one high density row and one low density row,
effectively inverting the potential experienced by
a portion of the vortices.
In general, systems that show a ratchet reversal due to collective
effects have an initial ratchet flow in the easy direction
of the substrate for low ac drive amplitudes
\cite{Villegas03,Lu07,Reichhardt05,PerezdeLara11,Reichhardt15}.
A similar type of ratchet reversal is studied
for superconducting vortex flows
in 2D Josephson networks where a similar
q1D asymmetric substrate potential arises
\cite{Marconi07}.
Studies of 2D active matter on 1D asymmetric substrates also show
that at low densities
the particle flow is in the easy direction,
but when collective effects come into play, there
can be a reversal
in which the flow is in the hard direction \cite{McDermott16}.
Ratchet reversals may also arise in systems
where inertial effects come into play \cite{Mateos00,Ai17}.

Another particle-based system is dusty plasma\cite{Chu94,Thomas94,Morfill09,Fortov05,Bonitz10,Merlino04,Melzer96,Nosenko04,Nunomura05,Knapek07}, or complex plasma,
where micron-sized dust particles
absorb free electrons and ions in plasma, so that these dust particles become highly charged to $\sim -{10}^{4} e$ typically in the steady state.
In dusty plasmas, the interaction between dust particles is the
Yukawa repulsion\cite{Konopka00,Ashwin15}, and
these particles are strongly coupled due to their high charges
\cite{Chu94,Thomas94,Morfill09,Wieben19}.
Under typical laboratory conditions, due to the electric field in the sheath, dust particles can self-organize into a single layer suspension, either in the crystal or liquid state, i.e., two-dimensional dusty plasma (2DDP) \cite{Morfill09,Hartmann13,Hartmann19}. In dust plasma experiments, individual particle identification and tracking from video imaging can be achived easily using high-speed cameras, due to the length scale of the interparticle distance and the time scale of the particle motion 
\cite{Morfill09,I96,Williams12,Thomas05}.
Dusty plasmas are typically used to study a variety of fundamental physics procedures in solids and liquids at the individual particle level, such as phase transitions
\cite{Nosenko09,Hartmann10,Singh22,Melzer12}, transport \cite{Nosenko04,Liu08,RomeroTalamas16,Hartmann19}, waves \cite{Nunomura05,Hartmann13,Piel08,Nunomura02a},
and other dynamics \cite{Lai02,Hartmann14,Hu22,Yu22}.
Unlike many other 2D strongly coupled
systems such as charged colloids in solution or
vortices in superconductors, the dust motion in the plasma environment is underdamped,
so that inertial effects
come into play, leading to
the appearance of phonons \cite{Nunomura02a,Couedel09}
and shock wave phenomena \cite{Samsonov99,Heinrich09,Kananovich21}.
The collective dynamics of dust particles can be modulated by applying substrates in experiments. For example, a stripe electrode is used to generate substrates to manipulate the transport of dust particles~\cite{Jiang09,Li10}. Recently, using simulations, the dynamics of dusty plasma modulated by periodic substrates are extensively investigated, such as the depinning dynamics of 2DDP modulated by 1D and 2D periodic substrates~\cite{Li19,Gu20,Huang22,Zhu22}.
Studying the dynamics of
dusty plasmas under substrates would
also be useful for understanding phenomena in
other particle systems where inertial
effects could come into play,
such as acoustically levitated particles \cite{Pandey19},
certain active matter systems \cite{Lowen20}, granular matter
\cite{Lim19},
and ions in trap arrays \cite{Bylinskii16}.
Due to its suitable length and temporal scales, dusty plasma is a natural model system to explore ratchet effects in the presence of both 
collective interactions and inertia.

In a recent experiment \cite{He20} performed in dusty plasma with a ring-shaped 1D dust particle chain confined inside an asymmetric-sawtooth-shaped channel, persistent flows of these dust particles are achieved. By changing the plasma conditions of the gas pressure or the plasma power, the direction of the dust particle flow is controlled. In combination with the corresponding numerical simulations, it is found that the asymmetric potential and collective effects are the two keys in the observed dusty plasma ratchet rectification and reversal. Other simulation results also show
dusty plasma ratchet effects in a q1D asymmetric potential
through measurements of the velocity distributions \cite{Chugh22}.

In this paper, using computer simulations, we investigate the collective transport of 2DDP under a one-dimensional asymmetric periodic substrate (1DAPS) driven by unbiased external excitations. The specified asymmetric periodic substrates are similar to the
geometries studied in superconducting vortex systems
\cite{Lu07,Shklovskij14,Marconi07}
or active matter systems \cite{McDermott16}, but different
from the geometry in Ref.~\cite{He20}. The rest of this paper is organized as follows.
In Sec.~II, we briefly introduce our simulation method to mimic 2DDP under ratchet substrates driven by unbiased external excitations. In Sec.~III, we present our simulation results in details, and also provide our interpretation of these results. We compare our results with the previous findings in 2D overdamped systems of superconducting vortices and active matter under similar 1D asymmetric potentials, and also discuss the possible future extension of the current work. 
Finally, we provide a brief summary of our findings in Sec.~IV.

\section{Simulation method}

We use Langevin dynamical simulations to investigate the collective transport properties of a 1DAPS modulated 2DDP driven by unbiased external excitation forces. The equation of motion for the dust particle $i$ is given by 
\begin{equation}\label{LDE}
{m \ddot{\bf r}_i = -\nabla \Sigma \phi_{ij} - \nu m\dot{\bf r}_i + \xi_i(t)+{\bf F}^{S}_i+{\bf F}^{D}_i. }
\end{equation}
Here, $-\nabla \Sigma \phi_{ij}$ is the particle-particle interaction force,
which has the form of a Yukawa or screened Coulomb potential $\phi_{ij} = Q^2 {\rm exp}(-r_{ij} / \lambda_D) / 4 \pi \epsilon_0 r_{ij}$, where $Q$ is the particle charge, $\epsilon_0$ is the permittivity of free space, $r_{ij}$ is the distance between particles $i$ and $j$, and $\lambda_D$ is the environment parameter of the Debye screening length due to the free elections and ions in plasma. The second term $- \nu m\dot{\bf r}_i$ is the frictional drag that is proportional to the particle velocity. The third term $\xi_i(t)$ is the Langevin random kicks, which are assumed to be Gaussian distributed with a mean of zero. According to the fluctuation-dissipation theorem~\cite{Feng08,vanGunsteren82}, $\langle \xi_i(0)\xi_i(t)\rangle = 2m\nu k_{B}T\delta(t)$, the magnitude $\xi_i(t)$ of the Langevin random kicks is related to the specified target temperature. The fourth term ${\bf F}^{S}_i$ is the force from the applied 1DAPS, which has the form  
\begin{equation}\label{1DAPSE}
{ U(x) = U_0[\sin(2 \pi x/w)+1/4\sin(4 \pi x/w)], }
\end{equation}
so that ${\bf F}^{S}_i = - \frac {\partial U(x)}{\partial x} \hat{\bf x} = (-\pi U_{0}/w)[2\cos(2\pi x/w)+\cos(4\pi x/w)] \hat{\bf x}$. Here, $U_{0}$ and $w$ are the depth and width of the 1DAPS, in units of $E_0 =Q^2/4\pi\epsilon_0 a$ and $a$, respectively, where $a = (n\pi)^{-1/2} $ is the Wigner-Seitz radius for the areal number density $n$. The last term on the right-hand side of Eq.~(\ref{LDE}), ${\bf F}^{D}_i = F_A \sin(\omega t) \hat{\bf x}$, is the applied unbiased sinusoidal excitation force, which has a mean value of zero.

Typically, 2DDPs, or 2D Yukawa systems, are described using two dimensionless parameters, which are the coupling parameter $\Gamma$ and the screening parameter $\kappa$~\cite{Morfill09,Bonitz10,Merlino04,Hartmann05,Kalman04}. These two dimensionless parameters are defined as $\Gamma = Q^2/(4 \pi \epsilon_0 a k_BT)$ and $\kappa = a / \lambda_D $, respectively, where $T$ is the dust particle kinetic temperature. From these definitions, the coupling parameter $\Gamma$ can be regarded as the inverse temperature, and the screening parameter $\kappa$ indicates the length scale of the space occupied by one dust particle, while the environment parameter of the Debye screening length $\lambda_D$ is assumed to be constant. Here, we fix $\Gamma = 1000$ and $\kappa = 2$ for our simulated 2DDP, which corresponds to the typical solid or crystal state without any substrates or external excitations~\cite{Hartmann05}. To normalize the length, in addition to the Wigner-Seitz radius $a$, we also use the lattice constant $b$, the average distance between nearest neighbors. For a 2D defect-free triangular lattice we have $b=1.9046a$. Time scales are normalized by the inverse nominal 2DDP frequency~\cite{Kalman04}, $\omega^{-1}_{pd} = (Q^2/2 \pi \epsilon_0 m a^3)^{-1/2}$, which is the typical time scale of interparticle collisions for strongly coupled 2DDP.

Our simulation includes 1024 particles within a rectangular box of dimensions $61.1a \times 52.9a$ with periodic boundary conditions. Since the simulated size in the $x$ direction is $61.1a \approx 32.07b$, in order to satisfy the periodic boundary conditions, we set the width of the potential well of the substrate $w$ to values that produce integer numbers of the potential well within the simulation box. Here, we set $w=2.004b$, corresponding to 16 full potential wells in the $x$ direction. For the depth of the 1DAPS, we specify four different values, $U_0/E_0 = 0.01$, $0.05$, $ 0.1$, and $ 0.2$, respectively. The expression of the external sinusoidal excitation force $F^{D}_i$ in Eq.~(\ref{LDE}) has two parameters, the amplitude $F_A$ and the angular frequency $\omega$, which are measured in units of $F_0=Q^2/4\pi\epsilon_0 a^2 $ and $\omega_{pd}$, respectively. We increase the amplitude of the sinusoidal excitation force $F_A$, and then measure the corresponding collective directional transport for different angular frequencies of $\omega/\omega_{pd} = 2\pi/140$, $2\pi/70$, and $2\pi/35$, respectively. The frictional damping coefficient is set to $\nu = 0.027\omega_{pd}$ in order to mimic the typical experimental conditions~\cite{Feng08a}. As in Ref.~\cite{Feng13}, we specify the cutoff radius of the Yukawa potential as $24.8a$. For each simulation run, we integrate Eq.~(\ref{LDE}) for $\ge 10^5$ steps with a time step of $0.0028\omega^{-1}_{pd}$ in order to reach a steady state, and then record the positions and velocities of all particles during the next several hundred periods of the sinusoidal excitation force for the data analysis reported here. Other simulation details are similar to those described in Ref.~\cite{Feng16a}. We also perform a few test runs for a much larger system containing 4096 particles to confirm no substantial differences from the results reported here.

\section{Results and Discussion}

\subsection{Structure of 2DDP on 1DAPS}

\begin{figure*}[htb]
	\centering
	\includegraphics{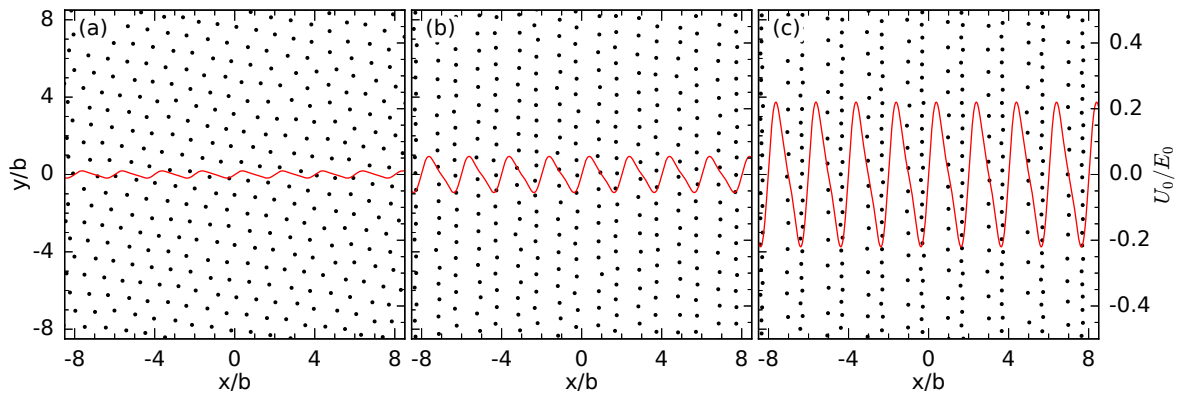}
	\caption{\label{fig:eps1} 
Snapshots of the particle positions (dots) of a 2DDP under the 1DAPS (curve) of $U(x)=U_0[\sin(2 \pi x/w)+1/4\sin(4 \pi x/w)] $ with various substrate depths $U_0/E_0$ of 0.01 (a), 0.05 (b), and 0.2 (c). The conditions of the simulated 2DDP are $\Gamma=1000$ and $\kappa=2$, corresponding to a solid state.
When $U_0/E_0 = 0.01$ in panel (a), the particles are uniformly distributed in a triangular lattice, independent of the locations of the potential wells of the 1DAPS. When $U_0/E_0 = 0.05$ in panel (b), two rows of particles sit in a stable zigzag arrangement within each potential well of the 1DAPS, forming a zigzag structure. 
When $U_0/E_0 = 0.2$ in panel (c), the two particle rows in each potential well become highly asymmetric, and the row at the bottom of the potential well contains many more particles than the other row. Note, only $\approx 32$\% of the total simulation box is plotted here. 
}
\label{fig:1}
\end{figure*}

We first focus on the structure or arrangement of the dust particles modulated by the applied 1DAPS with different depths of the potential well $U(x)$, in the absence of the excitation force.
Figure 1 presents snapshots of the simulated particle positions under the 1DAPS with different values of $U_0$, as well as sketches of the substrate potential. Clearly, the biharmonic function of $U(x)$ contains a single well or minimum in each spatial period, and the spatial asymmetry of the substrate $U(x)$ is present only in the $x$ direction. 
To the left side of the minimum
of the potential well the slope is gentle, while the right side is steep.
In Fig.~1, the parameters for the 2DDP and the spatial period of the 1DAPS are fixed to $\Gamma=1000$, $\kappa=2$, and $w/b=2.004$, while we vary only the substrate depth, as specified to $U_0/E_0 = 0.01$, $0.05$, and $0.2$ for Figs.~1(a), 1(b), and 1(c), respectively.

As shown in Fig.~1, when modulated by the applied 1DAPS with different depths, the simulated particles exhibit completely different arrangements. 
When the depth of the 1DAPS is shallow, such as $U_0/E_0 = 0.01$ in Fig.~1(a), the constraint from the substrate is negligible compared with the interparticle interaction, so that all particles form a nearly triangular floating solid lattice, independent of the locations of the potential wells of the 1DAPS.
When the substrate depth increases to $U_0/E_0 = 0.05$ as shown in Fig. 1(b), each potential well of the 1DAPS captures two ordered rows of particles that are aligned with the substrate potential. One row is at the minimum of the potential well, while the other is not, resulting in a
pinned smectic state, different from the arrangement of particles modulated by symmetric substrates~\cite{Gu20}.
Within each potential well, the particles adopt a stable zigzag arrangement to reduce the interparticle interactions.
For the deeper substrate of $U_0/E_0 = 0.2$ in Fig.~\ref{fig:1}(c), two distinctive rows of particles with different number densities appear in each 1DAPS potential well. The row on the steep side of the potential well contains many more particles than the other row, which is also closer to the potential minimum, due to the stronger trapping by the steep side of the substrate.
Clearly, as the depth of the 1DAPS becomes larger, the arrangement of particles exhibits a more pronounced asymmetry in the direction of the 1DAPS constraint.
This structural transition is caused by the competition between the particle-particle interactions and the particle-substrate interaction. The structures in Fig.~1 can be compared to the ordering of 2D superconducting
vortex assemblies under similar 1D
asymmetric substrates \cite{Lu07},
where for intermediate substrate depths there are
two rows of particles
in each potential minimum similar to Fig.~1(b),
while for large substrate depths
each minimum contains one dense row and one sparse row, just like Fig.~1(c).

\subsection{Bidirectional flow with varying depths of 1DAPS}

\begin{figure}[htb]
	\centering
	\includegraphics{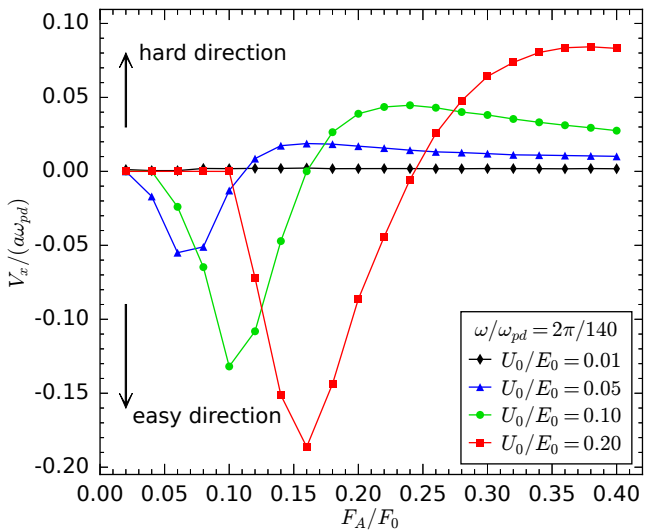}
	\caption{\label{fig:eps2} 
Calculated overall drift velocity $V_x$ of all simulated particles in a 2DDP with $\Gamma=1000$ and $\kappa=2$, as a function of the amplitude of the sinusoidal excitation force $F_A$, with the fixed angular frequency $\omega/\omega_{pd} = 2\pi/140$, for different substrate depths of $U_0/E_0 = 0.01$, 0.05, 0.10,
and 0.20. 
For the shallowest substrate of $U_0/E_0 = 0.01$, $V_x \approx 0$ for all values of $F_A$.
For $U_0/E_0 = 0.05$, 0.1, and 0.2, as $F_A$ increases gradually from zero, $V_x$ initially increases in magnitude with $F_A$ to a maximum value in the $-x$ direction, next decreases in magnitude back to zero, and then increases again to a maximum in the $+x$ direction before decaying gradually.
Thus, we find both a rectification of the particle motion and a reversal of the ratchet flow. The direction and magnitude of the overall drift flow velocity are both controlled by the depth of the 1DAPS and the amplitude of the unbiased sinusoidal excitation force.
}
\label{fig:2}
\end{figure}

To study the effect of the unbiased sinusoidal excitation force on the transport of a 1DAPS modulated 2DDP,
we calculate the overall drift velocity $V$ for all particles.
Since both the 1DAPS and the external sinusoidal 
excitation force are only in the $x$ direction,
the overall drift velocity in the $y$
direction should be always around zero under all studied conditions, which is confirmed in fact.
Thus, we focus only on $V_x$, the overall drift velocity in the $x$ direction, averaged over two hundred periods of the external sinusoidal excitation force using
\begin{equation}\label{CC}
{V_x = \langle N^{-1} \sum_{i=1}^N (\bm{v}_i \cdot \bm{\hat{x}}) \rangle,}
\end{equation}
as presented in Fig.~2. Clearly, from the specified 1DAPS in Fig.~1, the negative values of $V_x$ in Fig.~2 correspond to drift motion in the easy direction of the 1DAPS, while the positive values of $V_x$ indicate motion in the hard direction.
Here, we vary the amplitude and angular frequency of the external sinusoidal excitation force for various substrate depths $U_0/E_0 = 0.01$, 0.05, 0.1, and 0.2.

As the major result in this paper, we discover that the unbiased sinusoidal excitation force induces an overall drift flow of 2DDP in both the easy and hard directions of the 1DAPS, i.e., bidirectional flow. As presented in Fig.~2, the direction and magnitude of the overall drift velocity are
modified by changing the depth of
the 1DAPS and the amplitude of the unbiased external sinusoidal excitation force $F_A$.
For the shallowest substrate of $U_0/E_0 = 0.01$, the system forms a triangular lattice that floats
above the substrate, due to the tiny constraint from the 1DAPS. Thus, all particles move nearly rigidly back and forth under unbiased external excitations, resulting in a nearly zero net dc drift, as shown in Fig.~2.
From Fig.~2, when the substrate is deeper $U_0/E_0 \ge 0.05$, the particle arrangements are modulated by the 1DAPS, leading to a significant ratchet effect under the application of sinusoidal excitation forces.
As the amplitude of the sinusoidal excitation $F_A$ increases from zero, the overall drift velocity  $V_x$ first gradually increases from zero in the easy direction of the 1DAPS to its
maximum value, then gradually decreases in magnitude back to zero. As $F_A$ further increases, the drift velocity $V_x$ reverses toward the
steeper side of the 1DAPS, i.e., the hard direction, and increases to its maximum, then finally decays gradually when $F_A$ is large enough. 
From Fig.~2, clearly, as the substrate depth $U_0$ increases, the ratchet effect is more significant, with higher peak values in both the easy and hard directions.
In Fig.~2, it seems that, when the ratchet flow reversal occurs, the corresponding excitation amplitude $F_A$ also increases with the depth $U_0$ of the 1DAPS.

\subsection{Interpretation of bidirectional flow}

We attribute the observed ratchet effect in both the easy and hard directions to the combination of the spatial symmetry breaking of the 1DAPS and the inertial effects of the particles, as described in detail below.
In Fig.~2, the overall drift motion in the easy direction can be explained by the spatial symmetry breaking of the 1DAPS.
When the amplitude of the sinusoidal excitation $F_A$ is small, it is much easier for particles to move along the gentle side of each potential well than the steep side to cross the potential barrier of the substrate, leading to a diode-like effect. As a result, a higher probability for motion along the gentle side leads to an overall drift motion in the easy direction of the 1DAPS, i.e., the $-x$ direction.
As the excitation amplitude $F_A$ increases gradually, more particles are able to leap over the potential barrier of the substrate along the easy direction of the 1DAPS, naturally leading to an increase of the overall drift velocity.
In fact, if the excitation amplitude $F_A$ increases further, more particles are also able to go beyond the potential barrier along the hard direction as well, and as a result, the diode-like effect may be reduced. 
Within this range of the excitation amplitude $F_A$, clearly there is a peak in the magnitude of $V_x$ for motion in the $-x$ or easy direction, which occurs at $F_A / F_0 = 0.06$, 0.10, and 0.16 for $U_0 / E_0 = 0.05$, 0.10, and 0.20, respectively, as shown in Fig.~2.
When the excitation amplitude $F_A$ further increases, the overall drift velocity $V_x$ is still in the easy direction; however, its magnitude gradually decays until reaching zero.

If there were no inertia or other collective effects, this would be the end of the story and the ratchet
effect would disappear for higher values of $F_A$. Instead, however, we find a reversal
of the ratchet motion in the hard direction accompanied by
the appearance of a second peak in the magnitude of $V_x$ as $F_A$ increases further.

Indeed, the relative steepness of the decrease in $V_x$ from its negative
peak value in Fig.~2 already indicates that an additional mechanism has come into
play beyond the simple difference in the depinning threshold for motion in the
easy and hard directions. This mechanism is the emergence of a lag in the
response due to the presence of inertia in the dynamics of the simulated particles.
When the sinusoidal excitation force reverses its direction from the $+x$ to the $-x$ direction,
there is an overshoot in the motion of the particles due to the particle inertia, suggesting that the particles are unable to 
reverse their direction of motion immediately but instead continue moving in the $+x$ direction over
a specific distance $\delta d$ to reduce their velocity to zero, then begin to move in the $-x$ direction again.
Clearly, the magnitude of $\delta d$ increases with $F_A$ until
eventually it is the same size as the width of the steep side of the
potential. When this happens, particles are carried over the barrier in the
hard direction due to inertia alone. Since the distance from the
potential well minimum to the edge of the well in the easy flow
direction is longer, the inertia is insufficient to push the
particles over the barrier in the easy direction.
In essence, the presence of inertia causes a lag in the response
of the particles to the external sinusoidal excitation force,
resulting in a hysteretic behavior in the response of the particle velocity.
A steep tilt of the ratchet potential corresponds to a large hysteresis loop,
which results in a positive overall drift velocity. 

Now let us choose the condition of $U_0/E_0 = 0.10$ in Fig.~2 to explore the mechanism of the variation of $V_x$ with the amplitude of the sinusoidal excitation force $F_A$. For small amplitudes of the external sinusoidal excitation force,
such as $F_A/F_0 = 0.02$ or 0.04 for the 2DDP system with $U_0/E_0 = 0.10$,
the coupled particles are pinned at the bottom of the potential wells
and most cannot overcome the potential barriers,
leading to a collective overall drift velocity $V_x \approx 0$.
As the amplitude of the external sinusoidal excitation force
increases from $F_A/F_0 = 0.04$ to 0.1, more and more particles have enough energy to leap over the potential barrier in the easy direction of the 1DAPS, leading to an increase of the overall drift velocity in the easy direction.
When the amplitude of the external sinusoidal excitation force further increases from $F_A/F_0 = 0.1$ to $\approx 0.24$, the competition between
the directional probability and the
hysteresis due to inertial effects determines the overall net
transport direction of particles.
As the excitation force amplitude $F_A$ further increases, the directional probability effect decreases and the hysteresis effect increases, so a reversal of the overall drift motion occurs. As $F_A/F_0$ increases further well beyond 0.24, the amplitude of the external sinusoidal excitation force becomes large enough that the selective role of the ratchet potential drops slowly, so that the overall net collective drift velocity decreases gradually to zero.

\subsection{Diagnostics to support interpretation}

\begin{figure}[htb]
\centering
\includegraphics{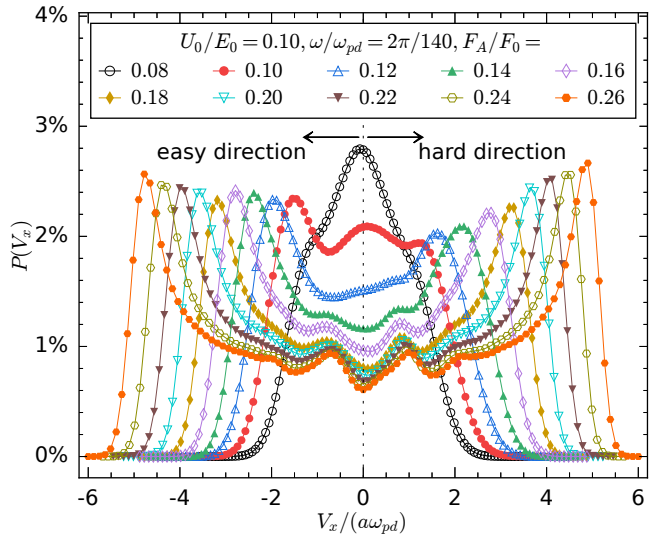}
\caption{\label{fig:eps3}
Calculated velocity distribution $P(V_x)$ of the particle motion for various excitation force amplitudes $F_A$ at fixed frequency $\omega/\omega_{pd} = 2\pi/140$ under the 1DAPS depth $U_0/E_0 = 0.1$. 
For the lower amplitudes $F_A/F_0 \le 0.14$, the probability of particles with negative values of $V_x$ is larger than that for positive values of $V_x$, indicating that particles are more likely to climb the potential barrier along the gentle side of 1DAPS, i.e., the easy direction.
For the higher amplitudes $F_A/F_0 \ge  0.18$, the height of the positive velocity peak in $P(V_x)$ is significantly higher than that of the negative velocity peak, indicating that the overall drift velocity changes to the hard direction.}
\label{fig:3}        
\end{figure}

To further confirm our interpretation above about the mechanism of the excitation induced drift flow in both the easy and hard directions of all particles under the 1DAPS, we analyze the velocity distribution $P(V_x)$. In Fig.~\ref{fig:3}, we plot our calculated $P(V_{x})$ versus $V_{x}$ under the condition of $U_{0}/E_{0} = 0.1$ for our studied system in Fig.~\ref{fig:2} with varying $F_{A}$ values. Clearly, when $F_{A}/F_0 = 0.08$,  $P(V_x)$ presented in Fig.~3 has a prominent peak at $V_x=0$, and the total distribution $P(V_x)$ of $V_x < 0$ is slightly greater than that of $V_x > 0$, resulting in a net drift velocity in the $-x$ direction. For $ F_{A}/F_0 = 0.1$, where the maximum magnitude of the easy direction ratcheting motion occurs, 
there are two peaks in $P(V_x)$ near $V_{x} = \pm 1.5 a\omega_{pd}$, but the peak for motion in the $-x$ direction is much higher than that for motion in the $+x$ direction, indicating that more particles climb along the easy direction of the 1DAPS than along the hard direction, naturally leading to an overall drift in the easy direction.
As $F_{A}$ further increases, the peak at $V_x=0$ diminishes
and velocity peaks appear only at finite positive and negative
velocities,
with the magnitude of the positive velocity peak increasing substantially with 
increasing $F_A$.
When $F_{A} / F_0 \ge  0.18$, the height of the positive velocity peak is significantly higher than that of the negative velocity peak, indicating that the overall drift velocity changes to the hard direction, agreeing well with the $V_x$ results in Fig.~2.
Clearly, there are also several smaller peaks near
$V_x = \pm a\omega_{pd}$, likely the result of
collective effects, that do not contribute much to the overall drift velocity $V_{x}$. 

\begin{figure}[htb]
\centering
\includegraphics{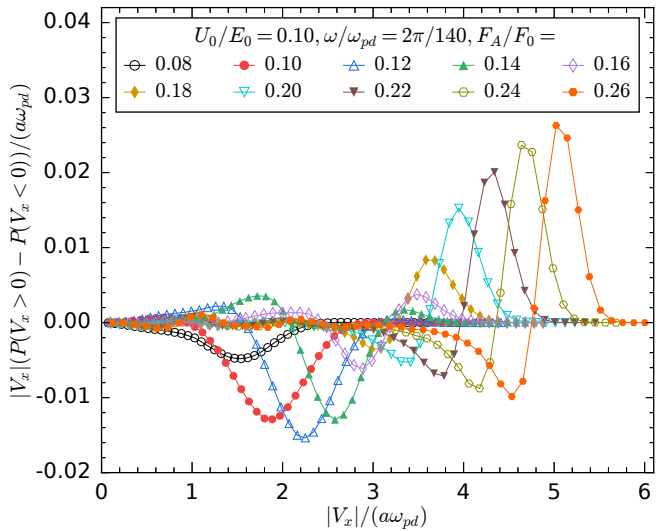}
\caption{\label{fig:eps4}
Calculated probability current of particle motion $|V_x|(P(V_x>0)-P(V_x<0))$ as a function of the speed $|V_x|$, for various excitation forces under fixed conditions of $U_0/E_0=0.10$ and $\omega/\omega_{pd}=2\pi/140$.
When the excitation force amplitude is small, $F_A/F_0 \le 0.12$, the negative peak of the probability current increases gradually with $F_A$, resulting in an increase in the overall drift velocity $V_x$ in the $-x$ direction, corresponding to a larger number of particles moving across the potential barrier in the easy direction.
When $ 0.14 \le F_A/F_0 \le 0.16$, the magnitude of the negative velocity peak of the probability current decreases gradually with $F_A$, while a peak at positive velocities appears and increases gradually with $F_A$, corresponding to more particles climbing the potential barrier along the steep side of the 1DAPS.
When $F_A/F_0 \ge 0.18$, the sinusoidal excitation force becomes large enough to drive all particles to climb the steep side of the 1DAPS, so that the positive velocity peak of the probability current increases significantly with the excitation amplitude $F_A$, leading to a net current in the $+x$ direction. This net flow direction is controlled by the asymmetry of the substrate and the amplitude of the unbiased excitation force.
}
\label{fig:4}
\end{figure}

In Fig.~\ref{fig:4} we present the calculated
probability current of particle motion as a function of the speed $|V_x|$ for various excitation forces with different amplitudes $F_A$ in our simulated 2DDP system with a ratchet substrate of $U_0/E_0=0.10$.
We obtain this quantity by multiplying $|V_x|$ by
$(P(V_x>0)-P(V_x<0))$ for each value of $|V_x|$.
For $F_{A}/F_0 \leq 0.12$, the net velocities
are mainly negative, while for $F_A/F_0=0.14$ to  $0.16$, the
system has a combination of both negative and positive velocities
that add up to a value close to zero.
When $F_{A}/F_0 \ge 0.18$, the probability current of the particle motion indicates that the net velocity is completely dominated by the substantial positive component of the velocity.

\begin{figure}[htb]
\centering
\includegraphics{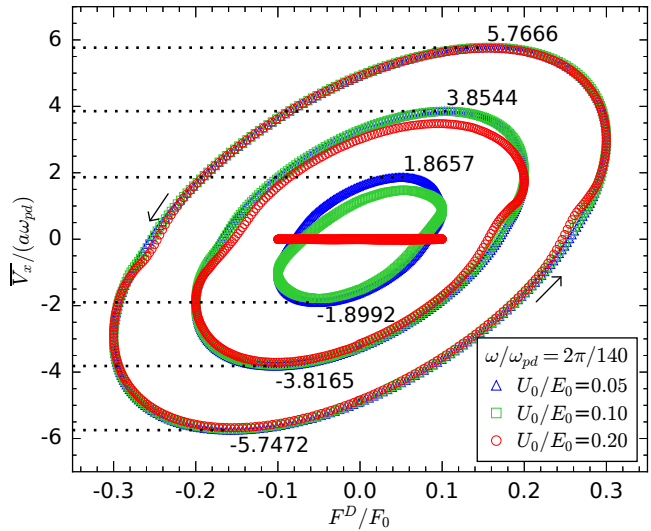}
\caption{\label{fig:eps5} 
Calculated instantaneous overall velocity $\overline{V_x}$ averaged over all particles, as a function of the instantaneous magnitude of the sinusoidal excitation $F^D$, during only one period of the excitation force, as $U_{0}/E_{0}$ varies. The significant hysteresis in the $\overline{V_x}$ versus $F^D$ plot clearly indicates an overshoot effect due to the inertia of the particles. From inner to outer sets of curves, $F_A/F_0=0.1$, 0.2, and 0.3.
}
\label{fig:5}
\end{figure}

In Fig.~\ref{fig:5} we plot the instantaneous velocity averaged over all particles $\overline{V_x}$ versus the instantaneous magnitude of the excitation force $F^D$ at different values of $F_A$, during only one period of the excitation force, for different 1DAPS depths of $U_0/E_0 = 0.05$, 0.10, and 0.20.
When $F_A/F_0=0.1$, under the condition of $U_0/E_0 = 0.2$, almost all particles remain pinned, so that $\overline{V_x}$ is nearly zero at all times. However, under the conditions of $U_0/E_0 = 0.05$ and 0.1, the amplitude of $F_A/F_0=0.1$ is large enough to induce depinning, so that particles are able to move across the 1DAPS.
For the conditions under which particles are able to move across the potential barrier of the 1DAPS, when the direction of the excitation force is reversed, the corresponding overall instantaneous velocity $\overline{V_x}$ for all particles is not able to respond immediately due to the inertia,
resulting in the appearance of hysteresis, also termed the overshoot effect~\cite{Gu20}.
For $F_A/F_0=0.1$,
the upper part of the cycle gradually increases in diameter
as the depth of the 1DAPS decreases,
indicating that more particles are able to cross the potential barrier on the steep side of the 1DAPS.
For $F_A/F_0=0.2$, the cycle at $U_0/E_0=0.20$ encloses a slightly smaller
area than the $U_0/E_0=0.05$ and $U_0/E_0=0.10$
cycles, suggesting that for $U_0/E_0=0.20$
some particles are not able to cross the barrier
on the steep side of the 1DAPS, leading to the maximum $\overline{V_x}$ in the $+x$ direction
below $3.8544 a\omega_{pd}$, which is the maximum $\overline{V_x}$ for $U_0/E_0=0.05$ or $U_0/E_0=0.10$.
For $F_A/F_0=0.3$, the extreme values of
$\overline{V_x}$ for
all the cycles at different depths of 1DAPS are nearly the same in the easy and hard directions, indicating that all of the particles are able to cross the potential barrier in both the $\pm x$ directions, leading to an overall net flow in the hard direction since the maximum peak in the hard direction of $\overline{V_x} = 5.7666 a\omega_{pd}$ is slightly larger.

Clearly, from our calculated velocity distribution $P(V_x)$ in Fig.~3, probability current $|V_x| (P(V_x>0)-P(V_x<0))$ in Fig.~4, and instantaneous overall velocity $\overline{V_x}$ versus $F^D$ in Fig.~5 presented above, we further confirm our interpretation of the observed ratchet effect for bidirectional flow of the particle motion. In our interpretation, the spatial symmetry breaking of the 1DAPS mainly induces a collective drift motion of particles in the easy direction of the 1DAPS when the amplitude of the excitation is small; however, when the amplitude of the excitation is large enough, the collective drift motion of particles in the hard direction is mainly caused by the combination of inertial effects of the particles and the spatial symmetry breaking of the 1DAPS. Our results presented above clearly indicate that the depth of the 1DAPS and the amplitude of the excitation can be used to modify the direction and magnitude of the overall drift flow. The next question is then whether the frequency of the excitation is also able to modify the overall drift velocity direction, as we study next.

\subsection{Bidirectional flow with varying frequencies of excitations}

\begin{figure}[htb]
\centering
\includegraphics{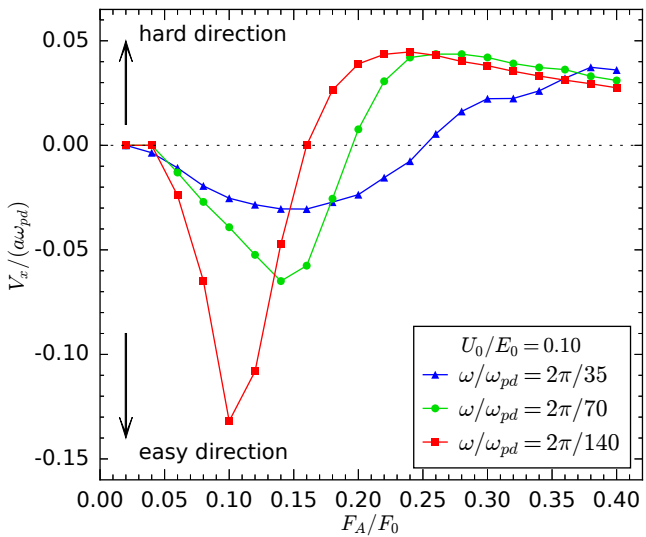}
\caption{\label{fig:eps6} 
		Calculated overall drift velocity $V_x$ of particles as a function of the excitation force amplitude $F_A$ with different angular frequencies $\omega/\omega_{pd} = 2\pi/140$, $2\pi/70$, and $2\pi/35$, under the condition of  unchanged substrate depth $U_0/E_0=0.1$. 
		For all three presented frequencies, as the amplitude $F_A$ of the excitation force gradually increases from zero, the overall drift velocity $V_x$ first increases from zero to its maximum in the $-x$ direction, next gradually decreases to zero, and then increases again to its another maximum in the $+x$ direction until it finally decays gradually.
		For one specific excitation force amplitude, the magnitude and the direction of the overall drift velocity $V_x$ are both determined by the excitation frequency $\omega$.
		That is to say, the flow rectification and its reversal can also be achieved by adjusting the frequency of the excitation force.
}
\label{fig:6}
\end{figure}

In Fig.~\ref{fig:6} we present our calculated overall drift velocity $V_x$ of our simulated 2DDP under an unbiased external sinusoidal excitation force with varying amplitudes $F_A$ and three different angular frequencies $\omega/\omega_{pd} = 2\pi/140$, $2\pi/70$, and $2\pi/35$, with the unchanged depth $U_0/E_0=0.1$ of the 1DAPS. 
For each angular frequency, as $F_A$ increases from zero, 
the overall drift velocity $V_x$ always first increases
in magnitude from zero to a maximum in the $-x$ direction, next gradually decreases back to zero, and then increases again to a maximum in the $+x$ direction before finally decaying gradually.
This variation trend of $V_x$ is very similar to that in Fig.~2.
Clearly, for one specified 1DAPS, the direction and magnitude of the overall drift velocity can be
adjusted by changing the amplitude $F_A$ and frequency $\omega$ of the excitation force.
From our interpretation, the underlying mechanism of the bidirectional flow in Fig.~6 should be the same as that in Fig.~2, which is the combination of the spatial symmetry breaking of the 1DAPS and the inertial effects of the particles.
From Fig.~6, as the frequency of the excitation force increases, the
maximum magnitude of the overall directional flow is reduced, probably because
the particles do not have enough time to respond to the variation of the excitation force. 
For a fixed excitation force amplitude,
the magnitude and direction of the overall drift velocity $V_x$ of
the particles are both determined by the excitation force frequency $\omega$.
In short, the flow rectification and its reversal of the overall drift velocity for all particles can be achieved by adjusting the excitation force frequency and amplitude.

\subsection{Discussion}

We would like to compare our results with the previous findings of superconducting vortices and active matter in 2D overdamped systems under similar 1D asymmetric potentials \cite{Lu07,McDermott16,Marconi07}. In such systems, at low densities the collective effects are minimal and there is only one row or a partial row of particles in each periodic potential minumum.
As an ac drive of increasing amplitude is applied, the system is first in a pinned phase. A transition occurs to finite dc flows in the easy direction of the substrate, while for larger ac drives, the dc flow drops back to zero. At higher densities, where two or more rows can fit in each potential minimum similar to what is shown in Fig.~\ref{fig:1}(b, c), the pinned phase is followed by finite dc flows in the hard direction of the substrate which reverse at higher ac drive amplitudes to a dc flow in the easy direction of the substrate.
The reversal results when there is a buckling
of the type illustrated in Fig.~\ref{fig:1}(c) in
the row of particles in each potential minimum,
so that a dense row and a sparse row coexist in a single substrate minimum. Under ac excitations, the particles in the sparse row are more mobile and experience both the substrate potential and the effective substrate potential induced by the repulsion from the dense row of particles.
Since the dense row is on the steep side of the potential minimum, it creates a steep barrier for the particles in the sparse row to jump out of the weaker side of the neighboring substrate potential.
In contrast, the
particles in the sparse row can push a portion of the particles in the dense row over the shorter but steeper side of the potential barrier, causing the particles in the dense row to perform a hop in the hard flow direction.
At higher ac drives, all of the particles can overcome the potential barrier, and the flow transitions to regular ratchet motion in the easy direction. 

Ratchet reversals are observed for other
collective ratchet systems
when the collective interactions lead to the emergence of
two effective species of particles.
The less strongly pinned species
experiences a substrate potential that is effectively inverted, and
this species undergoes ratcheting motion in the hard direction for lower
ac drives, while at higher ac drives, the noninverted potential
dominates and the ratcheting motion switches into the
easy direction \cite{deSouzaSilva06a,Gillijns07,Reichhardt05,PerezdeLara11}. 
This behavior is completely opposite from what we observe in Fig.~\ref{fig:2}. This suggests that 
although collective effects could still be occurring in the dusty plasma
system, the ratchet reversal is primarily due to the inertial effects.
For particles under a similar 1D asymmetric potential, it is
found that in the overdamped limit the ratchet flow
was only in the easy direction; however,
in an underdamped system
the ratchet motion occurs
in the easy direction for small ac drives but reverses
to the hard direction at higher ac drives \cite{Ai17},
similar to what we observe in Fig. 2.


The dusty plasma ratchet effect we study opens
several different future directions. 
Many other types of asymmetric potentials could be created
that may lead to ratchet effects, such as 2D asymmetric
sites \cite{deSouzaSilva06a,Gillijns07},
asymmetric confining walls \cite{Villegas03,Reichhardt05,PerezdeLara11},
and even systems
with frictional gradients \cite{Reichhardt15}. Since
the ratchet effect depends on
both the coupling strength and the inertia, it could be possible
to create dust sorting devices where particles of different mass move in
opposite directions or at different speeds.
If a magnetic field is applied to the dusty plasma \cite{Feng17},
gyroscopic effects may emerge
similar to those found for skyrmions,
and distinctive Magnus ratchet effects
could appear \cite{Reichhardt15a,Chen20,Gobel21}.
Since thermal effects are important,
it should also be possible to create a flashing ratchet where the
substrate is turned off and on to see whether there is directed motion
\cite{Reimann02,Rousselet94}. This would be most effective in the case
where the system is fluid like or near the solid to liquid transition.

\section{Summary}

In summary, we discover the bidirectional flow of a 1DAPS modulated 2DDP driven by unbiased sinusoidal excitation forces using Langevin dynamical simulations.
In the absence of the excitation force, the arrangement of the 2DDP is modulated by the 1DAPS. 
If the 1DAPS is shallow, the arrangement of the 2DDP is nearly unchanged from the typical triangular lattice.
When the depth of the 1DAPS is deeper, within each potential well, there are two rows of particles forming a zigzag structure. When the depth of the 1DAPS is further increased, the two
rows within one potential well become asymmetric, where one is dense and close to the bottom of the potential well, while the other is sparse and sitting on the gentle side of the potential well.
After applying a sinusoidal excitation force to the 1DAPS modulated 2DDP, we find a particle flow rectification and its reversal while changing the amplitude of the excitation force.
As the amplitude of the excitation force increases from zero,
the overall drift velocity of the 2DDP first increases
in magnitude from zero to a maximum in the easy direction of the 1DAPS, next gradually decreases back to zero, and then increases again to a maximum in the hard direction before finally decaying gradually.
The magnitude of the ratchet effect increases with the depth of the 1DAPS, and the maximum of
the ratchet effect shifts to higher values of the amplitude of the excitation force.
Furthermore, the frequency of the excitation force can be used to modify the magnitude of the ratchet effect and the location of the transition of the flow direction.
We attribute the observed ratchet effect in both the easy and hard directions to the combination of the spatial symmetry breaking of the 1DAPS and the inertial effects of the particles, as further confirmed by the three different presented diagnostics.

\section{Acknowledgments}

Work in China was supported by the National Natural Science Foundation of China under Grants Nos. 12105147 and 12175159, startup funds from Nantong University, and Innovation and Entrepreneurship Program of Jiangsu Province. Work at LANL was supported by the U. S. Department of Energy through the Los Alamos National Laboratory. Los Alamos National Laboratory is operated by Triad National Security, LLC, for the National Nuclear Security Administration of the U. S. Department of Energy (Contract No. 892333218NCA000001).


\end{document}